\documentstyle[12pt,epsfig,wrapfig]{article}

\begin{document}
\topmargin -12mm \oddsidemargin -4mm
\setcounter{page}{1}

\begin{center}
{\bf HARD POLARIZED PHOTON EMISSION IN SINGLE CRYSTALS BY HIGH ENERGY
ELECTRONS FOR PLANAR CRYSTAL ORIENTATIONS \footnote{{\bf The work is
supported by ISTC grant A-099}}\\[0pt]
\vspace{2mm} S.M.Darbinyan$^a$ and N.L.Ter-Isaakyan $^b$\\[0pt]
\vspace{2mm} {\small Yerevan Physics Institute Yerevan 375036, Armenian\\[0pt%
]
$^a$ E-mail: simon@lx2.yerphi.am\\[0pt]
$^b$ E-mail: terisaak@jerewan1.yerphi.am}\\[0pt]
}
\end{center}
\begin{abstract}
The radiation emission spectra of polarized photons emitted from charge
particle in single crystal are obtained in semiclassical theory in
Baer-Katkov-Strakhovenko approximation  for planar crystal
orientation. The range of applicability of this approximation is
estimated by comparing the results with calculations in exact
semiclassical theory. Optimal crystal orientations for producing unpolarized
and polarized photon beams are also founded.
\end{abstract}
{\bf 1. Introduction.} It is well known that the processes of photon
emission from charged particles (as well as $e^{-}e^{+}$ pair creation by
photon) at high energies in oriented single crystals reveal very strong
angular and energy dependence. The photon emission spectra in crystals in
their maxima could exceed the corresponding values for amorphous target 100
and more times and the radiation appears essentially polarized. Therefore,
these processes are widely applied in experimental physics for the
production of high energy polarized photon beams, as well as for the
analyses of photon polarization. For incident angles $\vartheta $ to crystal
axis/planes $\vartheta \gg \vartheta _{v}$, where $\vartheta _{v}$ is the
characteristic angle given by $\vartheta _{v}=U_{0}/m$ ($U_{0}$ is the scale
of axial/planar potential and $m$ the electron mass) these processes are
well described in the theory of coherent bremsstrahlung (CB) and pair
production (CPP) [1]. This theory is constructed in the framework of the
first Born approximation in crystal potential and fails at small angles, 
where these processes become of magnetic bremsstrahlung nature.

In the papers of V.N. Bayer V.M. Katkov and V.M. Strakhovenko has been
developed a general theory of basic QED processes in strong crystalline
fields (see i.e. new English edition of their book [2]). This theory is not
restricted by the first Born approximation and is based on the semiclassical
character of motion of ultrarelativistic particles in strong fields.
However, the numerical calculations in the exact theory face serious
mathematical difficulties and some approximation and modeling methods have
been developed. First numerical results of the radiation emission spectra in
the semiclassical theory were published in [3] (in the framework of
additional modeling assumption); exact calculations were presented in [4].
In Ref. [5] an analytical method of calculations in the framework of
semiclassical approach was developed and new numerical results were
presented. The radiation emission spectra of polarized photons in
semiclassical approach were calculated in [2,6] and first numerical results
were presented in our previous paper [6]. These papers demonstrate an
essential difference of exact spectra from the corresponding results of
coherent theory at small angles $\vartheta _{0}\le \vartheta _{v}$. There is
a good agreement between first experimental results at small angles [7,8] and
these calculations.

Polarization characteristics of radiation emission in single crystals in the
region of applicability of CB are well investigated in experiments (at the
energies of incident electrons up to 10 GeV) and the CB theory describes
 the experimental results satisfactory[1].

{\bf 2. Main results. } In this paper we have derived the polarized photon 
emission spectra in the framework of Bayer-Katkov-Strakhovenko approximation of 
paper [3] (further BKS approximation) which is valid at small incident electrons
angles to one of crystallographic planes. We also modified the nonpolarized
equations of [3] and present all results in the form very convenient for
numerical calculations. Our calculation algorithm allows to reduce the
computing time of all polarized spectra to 5-10 sec (Pentium 233) depending on
accuracy and crystal orientation.

Comparing the calculation in BKS approximation [4] with exact calculations
in semiclassical theory we estimated the range of applicability of this
approximation. It appears, that the equation in BKS approximation
 described the exact spectra satisfactory only at very small angles of
incidence, about 3-5 times less than critical angle of planar channeling.

We present a large number of polarized emission spectra in wide range of
angles for different energies and crystal orientations. The analysis of
these results make it possible to find optimal crystal orientations for
producing high energy photon beams. For producing nonpolarized high energy
photons it seems optimal to use the range of hard photon peaks at large $x$,
where $x$ is a fraction of electron energy $\varepsilon $ carried by emitted
photon. The maximum of intensity we find for $<110>$ oriented crystal for
incident electron moving close to $(1\bar{1}1)$ or $(1\bar{1}0)$ planes.
However, the degree of polarization at hard photon peaks appeared to be
rather low. We find that the degree of polarization at given $x$ achieves
its possible maximal value ($\sim 55\%$ at $x$=2/3, at $\varepsilon =200$
GeV) at relatively large values of incident angles to crystal axes, when the
corresponding hard peak situated at given  $x$ is of pure coherent
nature. The optimal configurations for producing polarized high energy
photons we find for{\it \ }$<001>$ oriented diamond crystal when electron
moves close to $(1\overline{1}0)$ plane and for $<110>$ oriented crystal when
electron also moves close to $(1\overline{1}0)$ plane.

{\bf 3. Polarized photon spectrum in BKS approximation.} Let us start from
the general semiclassical formula for polarized photon spectrum [2.6], which
we present in the following form in terms of Stocks parameters $\xi _{1}$, $%
\xi _{2}$, $\xi _{3}$:\newline
\begin{equation}
\frac{dI}{d\omega }=\frac{\alpha }{\pi }\frac{m^{2}\omega }{\varepsilon ^{2}}%
\int \frac{d^{3}r}{V}F({\bf {r}},\vartheta )\left[ \int_{0}^{\infty }\,\frac{%
d\tau }{\tau }[(T_{0}+\gamma ^{2}\xi _{1}T_{1}+\gamma ^{2}\xi _{3}T_{3})\sin
A_{1}+\gamma ^{2}\xi _{2}T_{2}\cos A_{1}]-\frac{\pi }{2}\right]  \label{AD}
\end{equation}
Here $\varepsilon $ and $\omega $ are electron and photon energies, $\gamma
=\varepsilon /m$; ${\bf {r}}$ is the entry point, $V$ is crystal volume,
integration over ${\bf {r}}$ gives the averaging over electron trajectories
in crystal, $F({\bf {r}},\vartheta )$ is the coordinate distribution
function at given value of $\vartheta $. Eq.1 must be also
averaged over time $t$. The quantities of $A_{1}=A_{1}(\tau )$ and $%
T_{i}=T_{i}(\tau ),i=0,1,2,3$, in (1) in rectilinear trajectory
approximation have the following form:

\begin{equation}
A_{1}=\frac{m^{2}\omega \tau }{\varepsilon \varepsilon ^{\prime }}\left[
1+\sum_{{\bf q,q}^{\prime }}\frac{G({\bf {q}})G({\bf {q^{\prime }}})}{%
m^{2}q_{z}q_{z}^{\prime }}{{\bf {q_{\perp }q_{\perp }^{\prime }}}}\left( 
\frac{\sin ((q_{z}+q_{z}^{\prime })\tau )}{(q_{z}+q_{z}^{\prime })\tau }-%
\frac{\sin (q_{z}\tau )\sin (q_{z}^{\prime }\tau )}{q_{z}q_{z}^{\prime }\tau
^{2}}\right) e^{-i\left( {\bf q+q}^{\prime }\right) {\bf r}(t)}\right] 
\label{AL}
\end{equation}
\begin{equation}
T_{0}=1-\varphi (\varepsilon )\sum_{{\bf {q},{q^{\prime }}}}\frac{G({\bf {q}}%
)G({\bf {q^{\prime }}})}{m^{2}q_{z}q_{z^{\prime }}}{\bf {q}}_{\perp }{\bf {q}%
}_{\perp }^{\prime }\sin (q_{z}\tau )\sin (q_{z}^{\prime }\tau )e^{-i\left( 
{\bf q+q}^{\prime }\right) {\bf r}(t)},  \label{AK}
\end{equation}
\begin{equation}
T_{1}=-\sum_{{\bf {q},{q}^{\prime }}}\frac{G({\bf {q}})G({\bf {q}}^{\prime })%
}{m^{2}q_{z}q_{z}^{\prime }}(q_{x}q_{y}^{\prime }+q_{y}q_{x}^{\prime
})[g(q_{z}\tau )g(q_{z}^{\prime }\tau )+\sin (q_{z}\tau )\sin (q_{z}^{\prime
}\tau )]e^{-i\left( {\bf q+q}^{\prime }\right) {\bf r}(t)},  \label{AO}
\end{equation}
\begin{equation}
T_{2}=\varphi (\varepsilon )\sum_{{\bf {q},{q}^{\prime }}}\frac{G({\bf {q}}%
)G({\bf {q}}^{\prime })}{m^{2}q_{z}q_{z}^{\prime }}(q_{x}q_{y}^{\prime
}-q_{y}q_{x}^{\prime })[g(q_{z}\tau )\sin (q_{z}^{\prime }\tau
)-g(q_{z}^{\prime }\tau )\sin (q_{z}\tau ]e^{-i\left( {\bf q+q}^{\prime
}\right) {\bf r}(t)},  \label{AM}
\end{equation}
\begin{equation}
T_{3}=-\sum_{{\bf {q},{q}^{\prime }}}\frac{G({\bf {q}}),G({\bf {q}}^{\prime
})}{m^{2}q_{z}q_{z}^{\prime }}(q_{x}q_{x}^{\prime }-q_{y}q_{y}^{\prime
})[g(q_{z}\tau )g(q_{z}^{\prime }\tau )+\sin (q_{z}\tau )\sin (q_{z}^{\prime
}\tau )]e^{-i\left( {\bf q+q}^{\prime }\right) {\bf r}(t)},  \label{AN}
\end{equation}
\[g(x)=\frac{\sin x}{x}-\cos x.\]
\noindent The notation used here coincides, in general, with ones of
Ref.[2]. The crystal potential is used in the form of $U(r)=\sum {G({\bf {q}%
})e^{-i{\bf {qr}}}}$, where {\bf {q}} is reciprocal lattice vector; $%
\varepsilon ^{\prime }=\varepsilon -\omega $, $\varphi (\varepsilon )=\frac{%
\varepsilon ^{\prime }}{\varepsilon }+\frac{\varepsilon }{\varepsilon
^{\prime }}$, $\,{\bf r}(t)={\bf r}+{\bf v}_{0}t$. The Stocks parameters and 
projections of vectors
correspond to the frame defined by following unit vectors: 
\begin{equation}
\hat{e}_{x}=[{\bf {n}}_{2}{\bf {v}}_{0}],\;\hat{e}_{y}={\bf {n}}_{2},\;\hat{e%
}_{z}={\bf {v}}_{0},  \label{AT}
\end{equation}
where ${\bf v_{0}}$ is the incoming electron velocity, ${\bf n_{2}}$ is a
unit vector transverse to ${\bf v_{0}}$ and to the main crystallographic
axis.

Let $\vartheta $ be the electron incident angle to the given
crystallographic axes and $\psi $ be electron incident angle to the given
crystallographic plane, then $\sin \psi =\sin \vartheta \sin \varphi $,
where $\varphi $ is the angle of ${\bf {v}}_{0}$ projection onto the plane
perpendicular to the given crystallographic axis, to the given
crystallographic plane.

The BKS approximation is valid for planar crystal orientations under the
following conditions.

\begin{enumerate}
\item  $\varphi \ll 1$, this makes it possible to isolate the continuous
planar potential contribution splitting the sums in (2-6) into two parts, $%
\sum_{{\bf q}}=\sum_{{\bf q}}^{F}+\sum_{{\bf q}}^{w}$, where the sum $\sum_{%
{\bf q}}^{F}$ contains ${\bf q}$ for which $q_{z}\rightarrow 0$ when $%
\varphi \rightarrow 0$ and the sum $\sum_{{\bf q}}^{w}$ includes all other
terms.

\item  $\psi \ll 1$, $\psi $ should be small enough\thinspace , so that $%
|q_{z}|\tau \ll 1$, this makes it possible to expand the functions in the sum 
$\sum_{{\bf q}}^{F}$ in powers of $|q_{z}|\tau $.

\item  $\vartheta $ should not be too small, $\vartheta \gg \vartheta _{v}$
so that the sum $\sum_{{\bf q}}^{w}$ be small enough and Eqs.2-6 could be
expanded in powers of $\sum_{{\bf q}}^{w}$.
\end{enumerate}
Two last conditions crucially depend on the effective upper integration
limit over $\tau $ in (1), $\tau _{0}$. The value of $\tau _{0}$ could not
be exactly established from general considerations due to rather complicated
dependence of Eq.2 on $\tau $, so, further we try to estimate the range of
validity of BKS approximation by comparing the BKS results with calculations
in exact semiclassical theory.

Following the steps of Ref.3, we did the above mentioned expansions, performed
time averaging and integrated the results over $\tau $. All polarized
intensity spectra we succeeded to present via one fold integrals of known
special functions: 
\begin{equation}
\frac{dI}{d\omega }=\frac{\alpha }{\pi }\frac{m^{2}\omega }{\varepsilon ^{2}}%
\int \frac{dy}{d_{pl}}F(y,\vartheta )\left( I_{0}+\xi _{1}I_{1}+\xi
_{2}I_{2}+\xi _{3}I_{3}-\frac{\pi }{2}\right) ,  \label{f}
\end{equation}
where the integration is taking over coordinate $y$ perpendicular to the
plane, $d_{pl}$ is the distance between the planes. Circular polarization
vanishes in BKS approximation, $I_{2}=0$. The quantities $I_{0},I_{1},I_{3}$
are defined as follows: 
\begin{eqnarray}
I_{0} &=&F_{0}(z,\lambda )+\beta \chi ^{2}F_{2}(z,\lambda )+  \nonumber \\
&&\sum^{w}\frac{G^{2}q_{\perp }^{2}}{m^{2}q_{z}^{2}}\ \left\{ \left( \frac{%
\beta }{4}+\frac{u}{s^{2}}(z-\frac{8u({\bf \chi q}_{\perp })^{2}}{%
s^{2}q_{\perp }^{2}})\right) \left( 2F_{0}(z,\lambda )-F_{0}(z_{+},\lambda
)-F_{0}(z_{-},\lambda )\right) \right. -  \nonumber \\
&&\frac{u}{s}\left( F_{0}(z_{+},\lambda )-F_{0}(z_{-},\lambda )\right) +\frac{%
u^{2}\chi ^{2}}{s^{2}}\left( 1-\frac{2({\bf \chi q}_{\perp })^{2}}{\chi
^{2}q_{\perp }^{2}}(1-\beta (1+\frac{\rho }{2})+\frac{4\beta \chi ^{2}}{s^{2}%
})\right) 2F_{2}(z,\lambda )-  \nonumber \\
&&\frac{u^{2}\chi ^{2}}{s^{2}}\left( 1+\frac{2({\bf \chi q}_{\perp })^{2}}{%
\chi ^{2}q_{\perp }^{2}}(1-\beta (1+\frac{\rho }{2})-\frac{4\beta \chi ^{2}}{%
s^{2}})\right) \left( F_{2}(z_{+},\lambda )+F_{2}(z_{-},\lambda )\right) - 
\nonumber \\
&&\frac{u\chi ^{2}\beta }{s^{2}}\left[ \left( 1-\frac{2({\bf \chi q}_{\perp
})^{2}}{\chi ^{2}q_{\perp }^{2}}\right) 2F_{1}\left( z,\lambda \right)
-\left( 1+\frac{2({\bf \chi q}_{\perp })^{2}}{\chi ^{2}q_{\perp }^{2}}%
\right) \left( F_{1}(z_{+},\lambda )+F_{1}(z_{-},\lambda )\right) \right] - 
\nonumber \\
&&\left. \frac{8u^{2}({\bf \chi q}_{\perp })^{2}}{s^{3}q_{\perp }^{2}}\left(
1-\beta (1+\frac{\rho }{2})\right) \left( F_{1}(z_{+},\lambda
)-F_{1}(z_{-},\lambda )\right) \right\}  \label{f0}
\end{eqnarray}

\begin{eqnarray}
I_{1} &=&\sum_{{\bf q}}^{w}\frac{G^{2}q_{x}q_{y}}{m^{2}q_{z}^{2}}\left\{ 
\frac{z^{2}}{s^{2}}\left( 2F_{0}(z,\lambda )-F_{0}(z_{+},\lambda )-F_{0}(z_{-},\lambda
)\right) \right. +  \nonumber  \label{f1} \\
&&\left. \frac{u\chi ^{2}}{s^{2}}\left[ 3\left( 2F_{1}(z,\lambda
)-F_{1}(z_{+},\lambda )-F_{1}(z_{-},\lambda )\right) +2zF_{2}(z,\lambda
)-z_{+}F_{2}(z_{+},\lambda )-z_{-}F_{2}(z_{-},\lambda )\right] \right\} 
\nonumber \\
&&
\end{eqnarray}

\begin{eqnarray}
I_{3} &=&\chi ^{2}F_{2}(z,\lambda )+  \nonumber \\
&&\sum_{{\bf q}}^{w}\frac{G^{2}q_{\perp }^{2}}{m^{2}q_{z}^{2}}\ \left\{ 
\frac{2u({\bf \chi q}_{\perp })^{2}}{s^{2}q_{\perp }^{2}}\left[
2F_{2}(z,\lambda )\left( z-\frac{4u\chi ^{2}}{s^{2}}\right) +\left(
F_{2}(z_{+},\lambda )+F_{2}(z_{-},\lambda )\right) \left( z+\frac{4u\chi ^{2}%
}{s^{2}}\right) \right] \right. -  \nonumber \\
&&\frac{u\chi ^{2}}{s^{2}}\left[ \left( 1-\frac{2({\bf \chi q}_{\perp })^{2}%
}{\chi ^{2}q_{\perp }^{2}}\right) 2F_{1}\left( z,\lambda \right) -\left( 1+%
\frac{2({\bf \chi q}_{\perp })^{2}}{\chi ^{2}q_{\perp }^{2}}\right) \left(
F_{1}(z_{+},\lambda )+F_{1}(z_{-},\lambda )\right) \right] +  \nonumber \\
&&\left. \frac{8u({\bf \chi q}_{\perp })^{2}z}{s^{3}q_{\perp }^{2}}\left(
F_{1}(z_{+},\lambda )-F_{1}(z_{-},\lambda )\right) \right\} +  \nonumber \\
&&\sum_{{\bf q}}^{w}\frac{G^{2}(q_{y}^{2}-q_{x}^{2})}{2m^{2}q_{z}^{2}}%
\left\{ \frac{z^{2}}{s^{2}}\left( 2F_{0}(z,\lambda )-F_{0}(z_{+},\lambda
)-F_{0}(z_{-},\lambda )\right) \right. -  \nonumber \\
&&\left. \frac{u\chi ^{2}}{s^{2}}\left[ 2F_{1}(z,\lambda
)-F_{1}(z_{+},\lambda )-F_{1}(z_{-},\lambda )-2zF_{2}(z,\lambda
)+z_{-}F_{2}(z_{+},\lambda )+z_{+}F_{2}(z_{-},\lambda )\right] \right\}
\label{f3}
\end{eqnarray}
where 
\begin{eqnarray}
{\bf \chi } &=&\frac{i\varepsilon }{m^{3}}\sum_{{\bf q}}^{F}G({\bf q}){\bf qe%
}^{-iq_{y}y},  \nonumber \\
\frac{\rho }{2} &=&\sum_{{\bf q}}^{w}\frac{\left| G({\bf q})\right|
^{2}q_{\perp }^{2}}{m^{2}q_{z}^{2}},  \label{ro}
\end{eqnarray}
$u=\omega /\varepsilon ^{\prime }$, \ $s=2\varepsilon q_{z}/m^{2}$, $%
\;\lambda ^{2}={\bf \chi }^{2}u$,\ $\beta =\varphi (\varepsilon )$,$%
\;\,z=u(1+\rho /2)$, $\;z_{\pm }=u(1+\rho /2\pm s/u)$. The functions $%
F_{0}(z,\lambda )$, $F_{1}(z,\lambda )$ and $F_{2}(z,\lambda )$ could be
expressed via known special functions. We present and analyze them in
{\bf Appendix}. The nonpolarized intensity spectra, $I_{0}$, coincide with result
of Ref.[2], if we take for $F_0(z,\lambda)$ the  integral representation 
via Bessel functions (Eq.7 of [2]).\\
\begin{wrapfigure}{r}{6.5cm}
\epsfig{figure=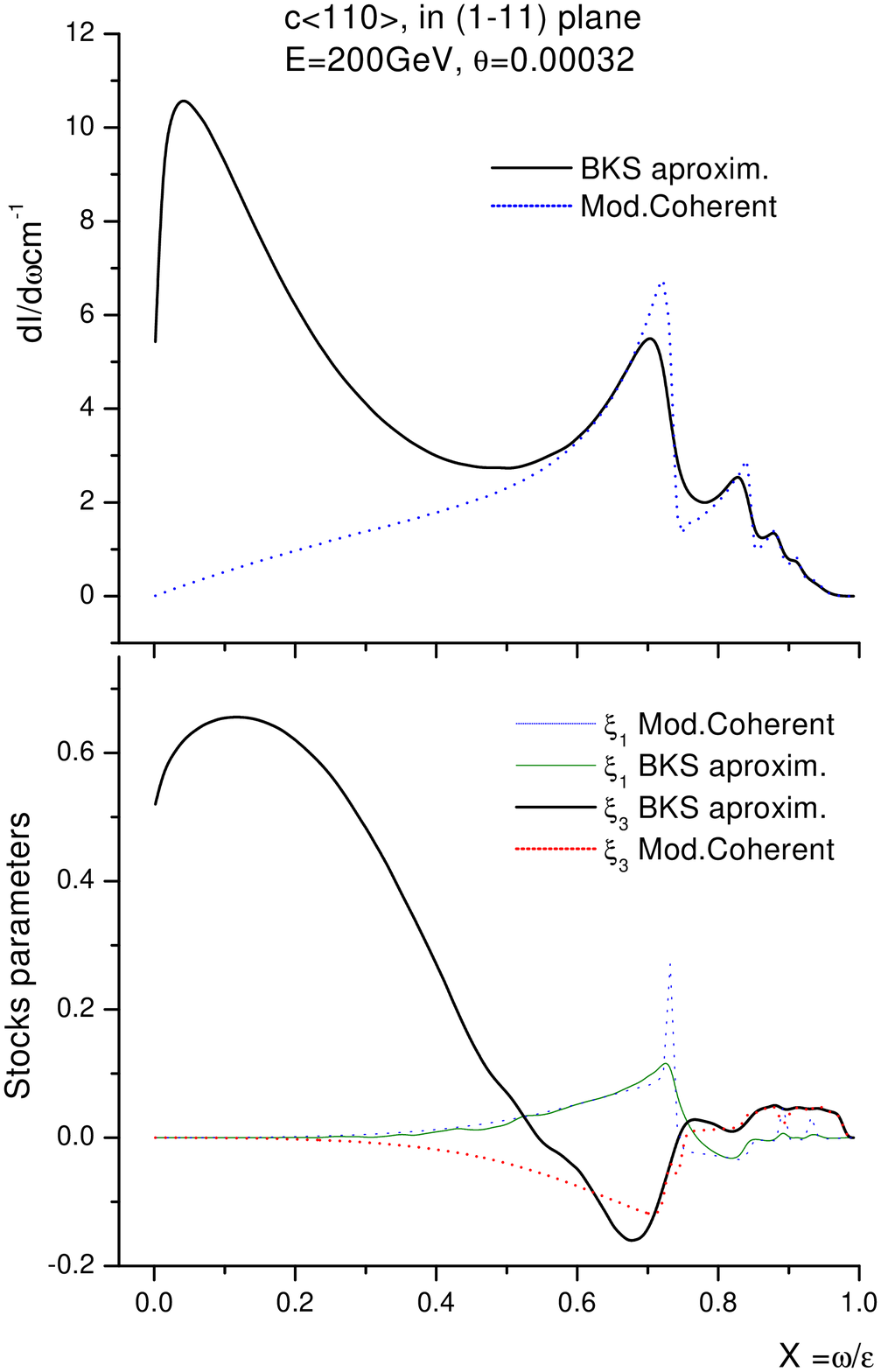,width=7cm}
{\small Fig.1: Comparison the hard part of spectral distributions of intensity 
and Stocks parameters in modified coherent and semiclassical (BKS) approaches}
\end{wrapfigure}
\indent {\bf 4.Numerical calculation results and conclusions}.In this section we
present the numerical calculations for uniform distributions, keeping,
of course, in mind, that for real spectrum one should take into account the
particle flux redistribution ($F(y,\theta _{0})\neq 1$) for $\psi \leq \psi
_{c}$, where $\psi _{c}$ is the critical angle of planar channeling, $\psi
_{c}=\sqrt{\frac{2U_{0}}{\varepsilon }}$, $U_{0}$ is the depth of planar
potential well. For $<001>$ orientation we define
the angle $\varphi $ as the angle of the projection of $v_{0}$ onto the
plane (001) to the plane $(1\overline{1}0)$, similarly, for $<110>$
orientation the angle $\varphi $ is the angle of the
projection of $v_{0}$ onto the plane (110) to the plane $(001)$. The typical
intensity spectrum and Stocks parameters are presented on Fig.1 
for $<110>$ oriented diamond crystal. Note, that throughout
the paper we present all angles in Radian.
The soft part of
the spectrum is completely determined by magnetic bremstrahlung
contribution. There is a wide polarization maximum of essential linear
polarization ($\xi _{1}=0,\;\xi _{3}\sim 0.5-0.7$) in wide range of $%
x\leq 0.3$. The shape of intensity and polarization spectra in soft region
are not changed by the variation of the angle $\varphi $, because in BKS
approximation the continuous planar potential contribution was actually
derived in the limit $\varphi \rightarrow 0$.
The high frequency peaks in hard part of the spectra could be
interpreted as coherent peaks slightly modified by influence of continuous
planar potential. This region can be approximately described by modified
coherent theory [2], if we exclude from the sums the reciprocal vectors
corresponding to magnetic bremstrahlung contribution ($q_{z}\rightarrow 0$
when $\varphi \rightarrow 0$). It is demonstrated on Fig.1. In such approximation 
each hard peak corresponds to the sum of coherent contributions of reciprocal vectors, 
so that for each peak corresponding values of $q_{z}$ tend to the fixed limit 
at $\varphi \rightarrow 0$ [6]. 
The polarization in hard region strongly depend on crystal orientation and will
be discussed later.

In order to estimate the range of applicability of BKS approximation we also
perform exact semiclassical calculations (making use of Eq.1-6) at different
incident angles to the plane\\
\epsfig{file=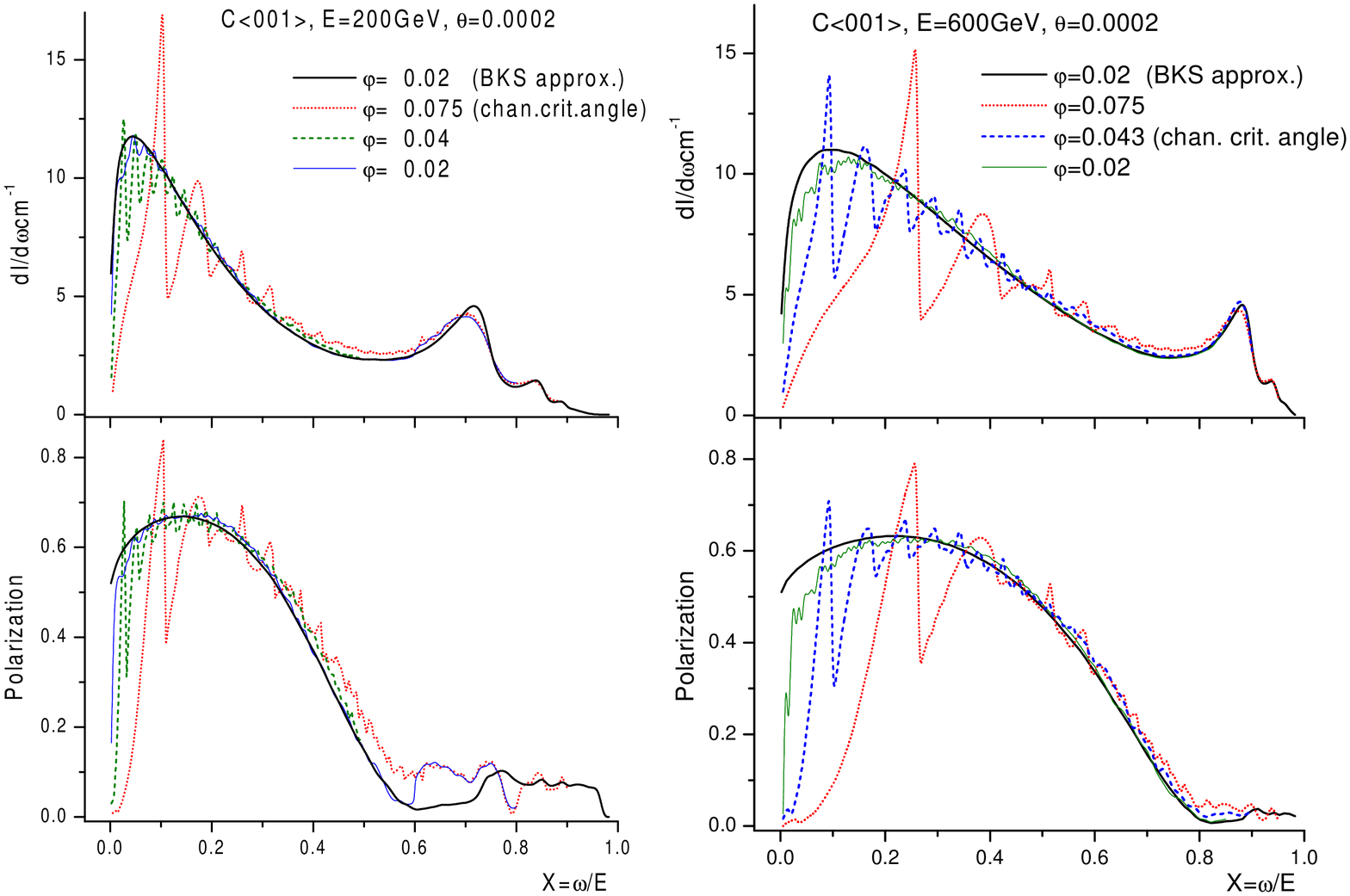,height=255pt,width=400pt,%
bbllx=0pt,bblly=10pt,bburx=800pt,bbury=580pt,%
rheight=243pt,rwidth=500pt}
{\small Fig.2: Comparison the spectral distributions of intensity and degree of
polarization of BKS approximation with exact semiclassical calculations.}

\vspace*{2mm}
\noindent for different energies. To carry out the
arising integrals over time of complicated oscillating function we have
elaborated special integration program. Results are presented in Fig.2. It
is very interesting that when electron incident angle to the plane is equal
or even less that planar channeling critical angle, $\psi _{c}$, the
soft part of the spectrum strongly differ from BKS \\
\epsfig{file=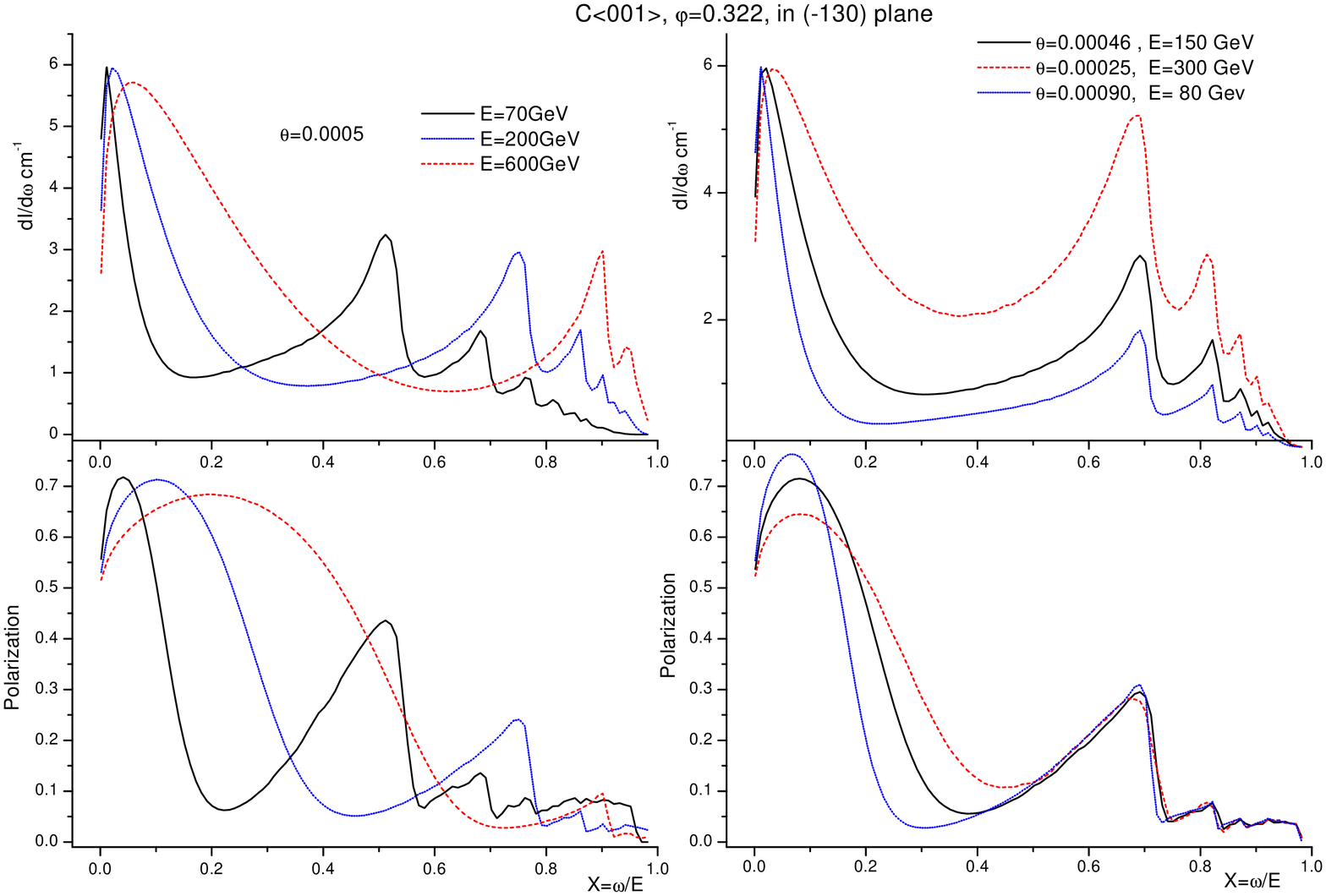,height=245pt,width=390pt,%
bbllx=0pt,bblly=10pt,bburx=800pt,bbury=580pt,%
rheight=237pt,rwidth=500pt}
{\small Fig.3: Energy dependence of spectral distributions of intensity 
and degree of polarization.}

\vspace*{2mm}
\noindent calculations and is far
from pure magnetic bremstrahlung nature. It more resembles the modified
coherent spectrum, with typical peaks and dips. That means that at these
values of incident angles the functions in the sums $\sum_{{\bf q}}^{F}$
cannot be expanded in powers of $|q_{z}|\tau $ in all integration region,
and together with magnetic bremstrahlung contribution, which comes from low $%
\tau $ region also contribute coherent effects which come from high $\tau $
region. Only at very small angles of incidence, about 3-5 times ( for 
$\varepsilon \sim 100-600$GeV) less than $\psi _{c}$, the soft
part of spectrum become of pure  magnetic bremstrahlung nature and could be
described in BKS approximation.
Note, that hard part of spectra is well described in BKS approximation
starting from large values of incident angles (practically from $\vartheta
\leq \vartheta _{v}$).\\
\epsfig{file=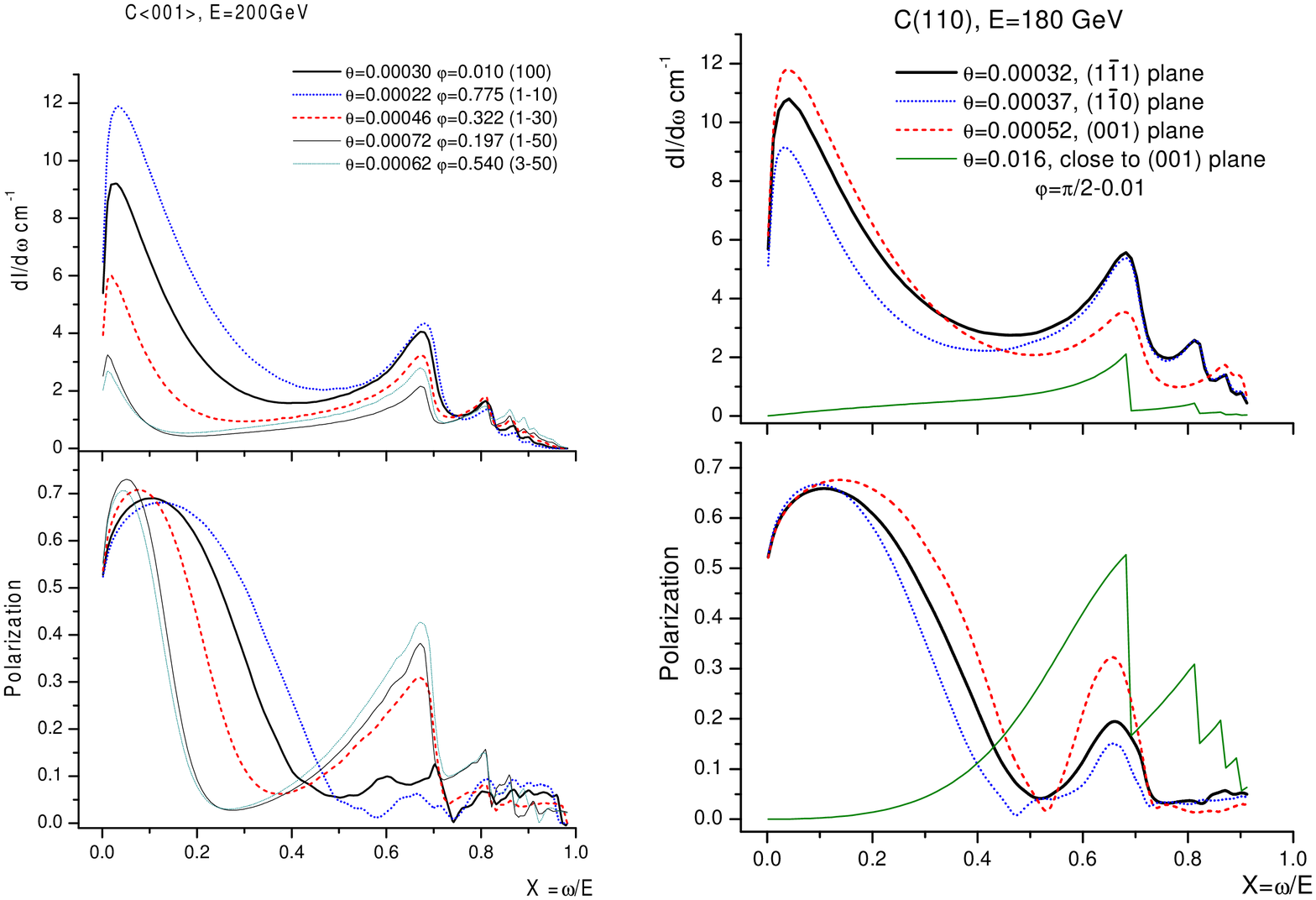,height=255pt,width=395pt,%
bbllx=0pt,bblly=10pt,bburx=800pt,bbury=580pt,%
rheight=245pt,rwidth=500pt}
{\small Fig.4: Spectral distributions of intensity 
and degree of polarization for different crystal orientations for fixed 
position of hard photon peak.}

\vspace*{3mm}
\noindent We also present large number of intensity and polarization spectra to
demonstrate the dependence of angles, energies and crystal orientations and
make it possible to find optimal configurations for producing high energy
photon beams. Some of the analogous nonpolarized intensity spectra were already presented
and analyzed in different papers [2-6]. The energy dependence at fixed
incident angles is demonstrated in Fig.3a. Hard peaks in intensity spectra
move to the left with increasing energy, slightly change their form, the
heights of the peaks being practically independent of energy, corresponding
peaks in polarization spectra also moved to the left and their heights
decrease. Soft peak also moves to the left and becomes wider. 
On Fig.3b we demonstrate the energy dependence of spectra at different angles, 
chosen in such a way that first hard peak is situated at fixed value of 
$x \sim 2/3 $. The height of hard peak in intensity spectra increase with 
increasing
energy, the hard part of polarization spectra practically remain unchanged.%
{\it \ So at very high energies it will be more and more preferable to use
high frequency peaks in photon spectra for producing high energy photon
beams}.
On Fig.4 we present the spectra for $<001>$ and $<110>$ diamond crystal
orientation when electron moves close to the main crystallographic planes at
fixed energy. Angles are arranged in such a way that first hard peak in all
spectra is situated at $x \sim 2/3$. {\it The maximal height of intensity
peak takes place for $<110>$ orientation when electron moves close to }$(1%
\overline{1}1)${\it \ or }$(1\overline{1}0)${\it \ planes and these
configurations are the most preferable for production of nonpolarized high
energy photons}. The degree of polarization in hard photon peaks for these
configurations appeared rather low. It is interesting that if we are going
from strong planes to weak ones, the intensity of hard photon peak is
decreasing and degree of polarization increasing. All these features could
be qualitatively understood in modified coherent theory. At hard photon
peaks the contributions of different planes are summed in nonpolarized
photon intensity ($I_{0}$), but in polarized photon intensities 
($I_{1}$, $I_{3}$) they could have opposite signs or one of the planes could not
contribute. Therefore, the degree of polarization very strongly depend on crystal
orientation. As one can see from Fig.4, for some orientations different contributions
in polarized intensities practically compensate each other. 
It seems impossible to find crystal orientation, for which the
contributions of all main plains will be summed in $I_{1}$ and $I_{3}$ .
Therefore, {\it the maximal polarization for hard photon peak at given value
of }$x${\it \ can be achieved for sufficiently large angles of incident
electron to the crystallographic axis when hard photon peak appeared to be
a typical coherent peak, corresponding to contribution of one plane and isn't
affected by influence of even very weak planes}. 
\begin{wrapfigure}{r}{6.8cm}
\epsfig{figure=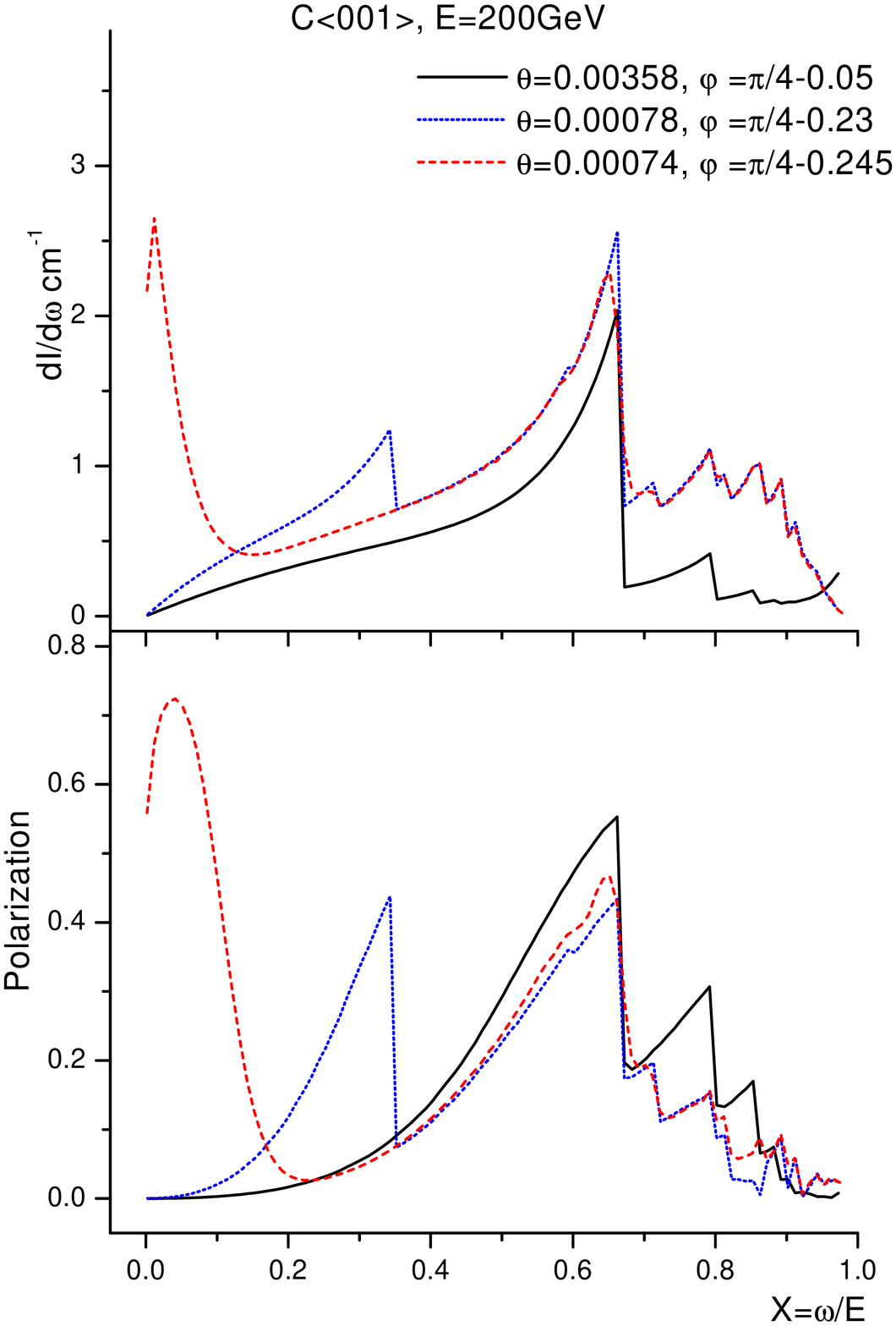,width=6.8cm}
{\small Fig.5: The influence of high order plane on intensity and 
polarization in coherent peak.}
\end{wrapfigure}
This is also
demonstrated in Fig.5 We present the intensity and polarization spectra for
$<001>$ diamond crystal for electron moving close to $(1\overline{1}0)$ plane
for large $\theta $ and different $\varphi $, keeping the position of main
peak fixed at $x \sim 2/3$. In wide range of $\varphi$ , starting from 
$\varphi \sim \pi /4$, height of the peak and degree of polarization
remain practically unchanged and coincide with presented curves for $\varphi$
=$\pi$/4-0.05, if we keep electron far from neighboring planes. When we come
closer to even rather weak planes, the height of the peak increases and the
degree of polarization decreases. We demonstrate this effect for weak 
$(\overline{3}50)$ plane ($U_{0} \sim 1.2$ eV). It is worth also noting
that when we come close enough to such week plane, we could see even
considerable magnetic bremstrahlung contribution. We have also checked this
properties for weaker planes, $(\overline{3}70)$ , $(\overline{5}70)$, 
$(\overline{3}90)$ and have found that their influence cannot be entirely
ignored.\\
\indent {\it The most preferable crystal orientations for production of polarized
high energy photons we find for }$<001>$ oriented diamond crystal when
electron moves close to $(1\overline{1}0)$ plane (for instance at $\theta
=0.028$ rad, $\varphi =0.735$ rad for $x \sim 2/3$) and for $<110>$
oriented crystal when electron also moves close to $(1\overline{1}0)$ plane
(for instance, at $\theta =0.028$ rad , $\varphi =1.536$ rad for $x \sim
2/3$). For these configurations the maximal degree of polarization could be
achieved for maximal intensities.These configurations bring exactly to the 
same results. The exact values of angles do not matter, one only should
arrange angles to fix the main coherent peak at given value of $x$ and keep
angles far from influence of neighboring planes.\\

\indent This work was supported by the ISTC grant, project A-099. One of us (N.T) 
is grateful to Prof. R.Avakian for suggestion to participate in this project.\\ 
\newpage

{\Large \bf Appendix}\\

\renewcommand{\theequation}{A.\arabic{equation}} \setcounter{equation}{0}

The function $F_{0}(z,\lambda )$ from Eqs.(9-11) can be expressed through
generalized hypergeometric functions, $_{p}F_{q}\left(
(a_{p});(b_{p});x\right) $: 
\begin{equation}
F_{0}(z,\lambda )=\frac{\pi }{6}+\frac{\pi z\;_{1}F_{2}\left( (\frac{1}{3});(%
\frac{2}{3},\frac{4}{3});\frac{z^{3}}{9\lambda ^{2}}\right) }{(3\lambda
)^{2/3}\Gamma (\frac{2}{3})}-\frac{3^{\frac{1}{6}}z^{2}\Gamma (\frac{2}{3}%
)\;_{1}F_{2}\left( (\frac{2}{3});(\frac{4}{3},\frac{5}{3});\frac{z^{3}}{%
9\lambda ^{2}}\right) }{4\lambda ^{4/3}}  \label{a}
\end{equation}

Functions $F_{1}(z,\lambda )$ and $F_{2}(z,\lambda )$ are expressed through
combinations of Bessel functions of the first kind, $J_{\nu }(x)$ , and of
the third kind, $K_{\nu }(x)$:

\begin{equation}
F_{1}(z,\lambda )=\theta (z)\frac{\sqrt{z}}{\sqrt{3}\lambda }K_{\frac{1}{3}%
}\left( \frac{2z^{3/2}}{3\lambda }\right) +\theta (-z)\frac{\sqrt{-z}}{%
3\lambda }\left( J_{\frac{1}{3}}\left( \frac{2(-z)^{3/2}}{3\lambda }\right)
+J_{-\frac{1}{3}}\left( \frac{2(-z)^{3/2}}{3\lambda }\right) \right)
\label{a1}
\end{equation}

\begin{equation}
F_{2}(z,\lambda )=\theta (z)\frac{z}{\sqrt{3}\lambda ^{2}}K_{\frac{2}{3}%
}\left( \frac{2z^{3/2}}{3\lambda }\right) +\theta (-z)\frac{z}{3\lambda ^{2}}%
\left( J_{\frac{2}{3}}\left( \frac{2(-z)^{3/2}}{3\lambda }\right) -J_{-\frac{%
2}{3}}\left( \frac{2(-z)^{3/2}}{3\lambda }\right) \right)  \label{a2}
\end{equation}

Here $\Gamma (x)$ is Euler gamma function and $\theta (x)$ is Heaviside
function. Note, that: 
\[
F_{1}(z,\lambda )=\frac{d\,F_{0}(z,\lambda )}{d\,z},\;\;F_{2}(z,\lambda )=-\frac{%
d\,F_{1}(z,\lambda )}{d\,z}. 
\]

Instead of making use of existing programs for calculations of special
functions, we have elaborated the calculation algorithm directly for
functions $F_{i}(z,\lambda )$, connecting the asymptotic expansion in powers
of $1/z$ with low $z$ expansion in power of $z$. We present the
corresponding equations only for $F_{0}(z,\lambda )$, analogous equations for $%
F_{2}(z,\lambda )$ and $F_{1}(z,\lambda )$ could be derived by
differentiation or from corresponding expansions of Bessel functions. The
asymptotic expansion of $F_{0}(z,\lambda )$ has the following form:

\begin{equation}
F_{0}(z,\lambda )=\frac{\pi }{2}-\frac{1}{2}\,\exp (-\frac{2z^{3/2}}{3\lambda })%
\sqrt{\frac{\pi \lambda }{z^{3/2}}}\left( 1-\frac{41\lambda }{48z^{3/2}}+%
\frac{9241\lambda ^{2}}{4608z^{3}}\right) ,\;\;z\gg 1  \label{as1}
\end{equation}
\begin{eqnarray}
F_{0}(z,\lambda ) &=&-\frac{\pi }{2}+\sqrt{\frac{\pi \lambda }{(-z)^{3/2}}}%
\left[ \left( 1+\frac{9241\lambda ^{2}}{4608z^{3}}\right) \cos \left( \frac{%
2(-z)^{3/2}}{3\lambda }+\frac{\pi }{4}\right) +\right.  \nonumber \\
&&\left. \frac{41\lambda }{48(-z)^{3/2}}\sin \left( \frac{2(-z)^{3/2}}{%
3\lambda }+\frac{\pi }{4}\right) \right] ,\hspace{3cm}z\ll -1  \label{as2}
\end{eqnarray}
Power series expansion in powers of  ~$z$~  follows from definition of
generalized hypergeometric functions: 
\begin{eqnarray}
F_{0}(z,\lambda ) &=&\frac{\pi }{6}+\frac{\pi \,z}{\left( 3\lambda \right)
^{2/3}\Gamma \left( \frac{2}{3}\right) }\left[ 1+\sum_{n=1}^{\infty }\frac{%
t^{n}}{n!}\prod_{k=1}^{n}\frac{3k-2}{(3k-1)(3k+1)}\right] -  \nonumber \\
&&\frac{3^{\frac{1}{6}\,}z^{2}\,\Gamma (\frac{2}{3})}{4\lambda ^{4/3}}\left[
1+\sum_{n=1}^{\infty }\frac{t^{n}}{n!}\prod_{k=1}^{n}\frac{3k-1}{(3k+1)(3k+2)%
}\right] ,  \label{a3}
\end{eqnarray}
where $t=z^{3}/3\lambda ^{2}$. Asymptotic expansions (A.4-A.5) work with
accuracy better than 0.1\% starting from $\left| z\right| /\lambda
^{2/3}\sim 4$ and Eqs. (A.4-A.5) and (A.6) could be connected at this
point. One should take at this point about 10 terms of expansion (A.6) .
Such calculation algorithm proved to be very effective and essentially
reduces the computer calculation time.


\begin{thebibliography}{99}
\bibitem{AR} M.L. Ter-Mikaelyan, High Energy Electromagnetic Processes on
Condensed Media, Wiley-Intercience, New-York,1972. 
\bibitem  V.N.Bayer, V.M. Katkov and V.M. Strakhovenko, 
Electromagnetic Processes at High Energies in Oriented Single Crystals,
World Scientific, Singapore, 1998.
\bibitem{AD}  V.N.Bayer, V.M. Katkov and V.M. Strakhavenko, Nucl.Instr.    
Methods. {\bf{B69}},(1992) 25.
\bibitem{BE} Yu.V.Kononets, I.S. Tupitsyn, Pisma Zh. Eksp. Teor. Fiz. 59,
491-497 (1994) 148-153.
\item V.M. Strakhovenko, Nucl.Instr. and Meth. {\bf{B145}}, 120 (1998).
\item S.M.Darbinyan, N.L.Ter-Isaakyan, hep-ph/9809559, 1998; 
JETP lett.{\bf 69},180, (1999). 
\end{thebibliography}
\end{document}